\documentclass[a4paper,11pt]{article}

\usepackage{amsmath}
\usepackage{amssymb}
\usepackage{graphicx}
\usepackage{hyperref}
\usepackage[a4paper,left=0.99in, right=0.99in,top=1.1in, bottom=1.2in]{geometry}
\usepackage{color}
\usepackage{cite}
\numberwithin{equation}{section}

\usepackage{epsfig}
\usepackage{graphicx,color}

\newcommand{\abel}{\text{(ab)}}
\newcommand{\noabel}{\text{(nab)}}
\newcommand{\be}{\begin{equation}}
\newcommand{\ee}{\end{equation}}

\usepackage{hyperref}

\usepackage{color}
\usepackage[normalem]{ulem}
\definecolor{pkcolor}{rgb}{0,0,1}
\definecolor{avhcolor}{rgb}{1,0,0}

\newcommand\pkout{\bgroup\markoverwith{\color{pkcolor}{\rule[0.4ex]{2pt}{0.8pt}}}\ULon}
\newcommand\avhout{\bgroup\markoverwith{\color{avhcolor}{\rule[0.4ex]{2pt}{0.8pt}}}\ULon}

\title{ 
    \vskip 2cm
    Saturation effects in forward-forward dijet production  \\
    in p+Pb collisions
}

\author{
A. van Hameren,$^1$ P. Kotko,$^1$ K. Kutak,$^1$ C. Marquet$^2$ and S. Sapeta$^3$\\\\
$^1$ {\small\it The H.\ Niewodnicza\'nski Institute of Nuclear Physics PAN}\\ {\small\it Radzikowskiego 152, 31-342 Krak\'ow, Poland}\\\\
$^2$ {\small\it Centre de Physique Th\'eorique, \'Ecole Polytechnique,}\\
{\small\it CNRS, 91128 Palaiseau, France}\\\\
$^3$ {\small\it CERN PH-TH, CH-1211, Geneva 23, Switzerland}\\
}

\date{}

\begin{document}
\maketitle

\vspace{-30em}
\begin{flushright}
  CERN-PH-TH/2014-029 \\
  CPHT-RR005.0214 \\
  IFJPAN-IV-2014-2
\end{flushright}
\vspace{25em}

\abstract{
We study saturation effects in the production of forward dijets in proton-lead collisions
at the Large Hadron Collider, using the framework of High Energy Factorization. Such
configurations, with both jets produced in the forward direction, probe the gluon density
of the lead nucleus at small longitudinal momentum fraction, and also limit the phase
space for emissions of additional jets. We find significant suppression of the forward
dijet azimuthal correlations in proton-lead versus proton-proton collisions, which we
attribute to stronger saturation of the gluon density in the nucleus than in the proton.
In order to minimize model dependence of our predictions, we use two different
extensions of the Balitsky-Kovchegov equation for evolution of the gluon density with sub-leading
corrections.
}

\section{Introduction}

The production of hadronic final states at the Large Hadron Collider (LHC) offers
unprecedented opportunities to test parton densities in various kinematic regions.
Of particular interest is the forward region, where it is possible to  construct
hadronic observables that allow to probe parton densities of one of the colliding
hadrons at longitudinal fractions $x\sim 10^{-5}$. At such low values of $x$, on
theoretical grounds, one expects ``low-$x$'' effects to be relevant, in particular the
phenomenon of gluon saturation \cite{Gribov:1984tu}.

In QCD, saturation is described by non-linear evolution equations for the gluon density, which
resum a subset of diagrams generating contributions of the form $\alpha_s\ln 1/x$. The solutions
of small-$x$ evolution equations, together with suitable initial conditions, provide parton densities,
which then need to be convoluted with appropriate hard matrix elements in order to obtain
predictions for measurable cross sections. The Color Glass Condensate~(CGC) (see \cite{Gelis:2010nm} and references therein) and High Energy
Factorization (HEF) \cite{Catani:1990eg} are two QCD-based frameworks which can be used for phenomenological studies.

The CGC approach has been very successful in describing forward di-hadron production at RHIC
\cite{Albacete:2010pg,Stasto:2011ru,Lappi:2012nh}, in particular it predicted the suppression of
azimuthal correlations in d+Au collisions compared to p+p collisions \cite{Marquet:2007vb}, which
was observed later experimentally~\cite{Adare:2011sc,Braidot:2010zh}. It has now become necessary
to extend the validity of the CGC predictions from RHIC kinematics to the LHC. There, the relevant
observables involve high-$p_t$ jets, as opposed to individual hadrons with $p_t$ of the order of a few
GeV at RHIC. Furthermore the advantage of jet observables as compared to hadrons is that they are less sensitive to large
uncertainties from fragmentation functions.

In order to accommodate this, the theoretical basis developed in the context of RHIC
collisions to compute di-hadron correlations must be supplemented with further QCD dynamics,
relevant at high $p_t$. This includes, for instance, coherence in the QCD evolution of the gluon density.
However, after accounting for higher-order corrections, it was recently argued that the CGC approach
to forward particle production may, at the moment, be only under control at low transverse momentum
$p_t\leq Q_s$, where $Q_s$ denotes the saturation scale, which precisely signals the onset of
parton saturation. There are conflicting results on this matter \cite{Stasto:2013cha,Kang:2014lha}, and
before this formulation can be used to perform forward jet studies at the LHC, further progress on the
theoretical side are needed to clarify the situation.

In view of these potential limitations, we shall instead investigate forward dijets at the LHC using a
different, more practical HEF framework. That framework was recently used to study the forward-central
dijet \cite{Deak:2009xt,Hautmann:2009kd} configuration \cite{Deak:2010gk,Deak:2011ga,CMS:2011xra,Kutak:2012rf}
and trijet production for forward-central and purely forward configuration \cite{vanHameren:2013fla}.
In this paper we employ it to study the forward-forward configuration, which offers some practical
advantages. First, that configuration is less demanding theoretically than the central-forward one
since the phase space for production of an additional third jet becomes limited. In addition, lower
values of $x$ can be accessed, enhancing the sensitivity to saturation effects.

In the context of the LHC, we will consider forward dijet production in p+Pb and p+p collisions.
Comparisons of p+Pb and p+p cross sections for the same observables can provide some evidences
for parton saturation since such effects are further enhanced by increasing the atomic number of one of
the colliding particles. In order to make an extensive study of saturation effects in forward-forward dijet systems,
we use gluon densities obtained from two different extensions of the original Balitsky-Kovchegov
equation \cite{Balitsky:1995ub,Kovchegov:1999yj}. The first one incorporates the running of the QCD
coupling \cite{Kovchegov:2006vj,Balitsky:2006wa}, and the second one also includes non-singular
pieces (at low $x$) of the DGLAP splitting function, a sea-quark contribution, and resums dominant
corrections from higher orders via a kinematic constraint \cite{Kwiecinski:1997ee, Kutak:2003bd}.

The paper is organized as follows. In section 2, we review the similarities and differences between the
CGC and HEF approaches to forward particle production. In section 3, we introduce the two evolution scenarios
that we shall consider for the small-$x$ non-linear evolution of the gluon distribution. In section IV, we present
our results for forward dijet production in p+p collisions at the LHC, as well as nuclear modification factors 
$R_{\rm pA}$ for p+Pb collisions. Finally, section V is devoted to summary.

\section{Color Glass Condensate vs High-Energy Factorization}

In this section, we recall the CGC description of single- and double-inclusive forward particle production in p+p (and p+A) collisions.
We outline the present limitations of that formalism when applied to high-$p_t$ jets, and introduce the HEF framework as a viable
practical alternative to obtain LHC predictions that include saturation effects.

\subsection{Forward particle production in the CGC}

In p+A (or p+p) collisions, particle production at forward rapidities is sensitive to large-$x$ partons from the proton, while the target nucleus (or the other proton) is probed deep in the small-$x$ regime. To compute cross sections in such an asymmetric situation, the appropriate formulation is the so-called hybrid factorization \cite{Dumitru:2005gt}, rather than the symmetric $k_{t}$-factorization adequate for mid-rapidity observables.

In the hybrid formalism, the large-$x$ partons are described in terms of the usual parton distribution functions of collinear factorization $f_{i/p}$, with a scale dependence given by DGLAP evolution equations, while the small-$x$ gluons of the nucleus are described by a transverse momentum dependent distributions, which evolve towards small $x$ according to non-linear equations. At leading order, single-inclusive hadron production is given by the following convolution of parton level cross sections with fragmentation functions $D_{h/i}$:
\begin{eqnarray}
\frac{d\sigma^{pA\to hX}}{dyd^2p_t}=\int_{x_F}^1\frac{dz}{z^2}\
\left[\sum_{q}x_1f_{q/p}
(x_1,\mu^2)\tilde{N}_F\left(x_2,\frac{p_t}{z}\right)D_{h/q}(z,Q^{2}) \right.\nonumber\\ +\left. 
x_1f_{g/p}(x_1,\mu^2)\tilde{N}_A\left(x_2,\frac{p_t}{z}\right)D_{h/g}(z,Q^{2})\right] 
\label{hybel}\ ,
\end{eqnarray}
where $x_1=x_F/z$ and $x_2=x_1e^{-2y}$ with $x_F=p_t e^y/\sqrt{s}$. The unintegrated gluon distributions $\tilde{N}_{F,A}$ are obtained from the dipole cross section by Fourier transformation:
\begin{equation}
\tilde{N}_{F(A)}(x,k)=\int d^2b \int \frac{d^2r}{(2\pi)^2}\
e^{-i{\bf k}\cdot{\bf r}}\left[1-N_{F(A)}(x,r,{\bf b})\right]\ ,
\label{phihyb}
\end{equation}
where $N_{F(A)}(x,r,{\bf b})$ is the imaginary part of the scattering amplitude of a fundamental (or adjoint) dipole of transverse size $r$ off the nucleus, at impact parameter ${\bf b}$.

This leading-order formula underlies a $2\to1$ partonic sub-process, and therefore cannot be matched onto standard perturbative results at high $p_t$. In order to accomplish that, next-to-leading order (NLO) corrections based on $2\to2$ kinematics are needed. Formally sub-leading in the saturation region, some of these corrections become leading in the high-$p_t$ regime, as explained in \cite{Altinoluk:2011qy}, where it was first realized that NLO corrections are crucial to bring the hybrid formulation of particle production into agreement with the standard perturbative result at large transverse momentum.

Recently, the full NLO corrections to the hybrid formalism have been calculated \cite{Chirilli:2011km,Chirilli:2012jd}. A first implementation indicates that these corrections are negative at high $p_t$, and in fact dominate over the leading-order result (\ref{hybel}), leading to a negative cross section \cite{Stasto:2013cha}. This suggests that calculations beyond NLO accuracy, or performing additional resummations at high $p_t$, are needed in order to stabilize the perturbative series. By contrast, an alternative work suggests a much simpler solution \cite{Kang:2014lha}. It may take some time until a consensus is reached, about whether or not the hybrid formalism as it is can be used to perform forward jet studies at the LHC, or if it is only under control at low transverse momentum $p_t\leq Q_s$.

Interestingly enough, there exist an alternative high-$p_t$ observable which is sensitive
to the saturation regime, where the theoretical formulation is under control: nearly back-to-back
dijets with a small transverse momentum imbalance $|{\bf p_{t1}}+{\bf
p_{t2}}|\sim Q_s\ll|{\bf p_{t1}}|,|{\bf p_{t2}}|$. While the general formulation
of double-inclusive particle production in the CGC is rather complicated, in
this nearly back-to-back situation the following factorization formula can be
derived within the hybrid formalism (in the large-$N_c$ limit)
\cite{Dominguez:2010xd,Dominguez:2011wm}:
\begin{eqnarray}
\frac{d\sigma^{pA\rightarrow {\rm dijets}+X}}{dy_1dy_2d^2p_{1t}d^2p_{2t}} =
\frac{\alpha _{s}^{2}}{(x_1x_2S)^{2}}\left[
\sum_{q}x_1f_{q/p}(x_1,\mu^2) \sum_i H_{qg}^{(i)} \mathcal{F}_{qg}^{(i)}(x_2,|{\bf
p_{1t}}+{\bf p_{2t}}|) \right.\nonumber\\ \left. 
+\frac12 x_1f_{g/p}(x_1,\mu^2) \sum_i H_{gg}^{(i)}
\mathcal{F}_{gg}^{(i)}(x_2,|{\bf p_{1t}}+{\bf p_{2t}}|)\right] \ ,
\label{dijet}
\end{eqnarray}
where the fractions of longitudinal momenta of initial state partons are related to
the transverse momenta $p_{1,2t}$ and rapidities $y_{1,2}$ of the final state partons by
\begin{equation}
  x_1 = \frac{1}{\sqrt{S}} \left(p_{1t} e^{y_1} + p_{2t} e^{y_2}\right)\ ,
  \qquad
  x_2 = \frac{1}{\sqrt{S}} \left(p_{1t} e^{-y_1} + p_{2t} e^{-y_2}\right)\,,
\end{equation}
and $\sqrt{S}$ is the center of mass energy of the p+A system.

$\mathcal{F}^{(i)}$ and $H^{(i)}$ are various unintegrated gluon distributions
and associated hard coefficients, respectively. Their expressions can be found
in Ref.~\cite{Dominguez:2010xd,Dominguez:2011wm}. In particular, one notes that the different
gluon distributions are expressed in terms of only two independent basic ones,
the so-called dipole gluon distribution (proportional to $\tilde{N}_F$), and the
Weizs\"acker-Williams gluon distribution related by Fourier transformation to a
quadrupole amplitude, and in some approximation to $\tilde{N}_A$. Even though we have in
mind the high-$p_t$ jets, we expect that the limitations of this hybrid formalism encountered in the
single inclusive case do not have an impact for dijets as long as the imbalance of the system
$|{\bf p_{t1}}+{\bf p_{t2}}|$ does not become larger than $Q_s$, since this is the argument
entering the gluon distributions $\mathcal{F}^{(i)}$.

It is a numerically difficult work to solve the quadrupole evolution equation. Instead, using models
for these gluon distributions as opposed to actual QCD evolution equations, formula (\ref{dijet})
was successfully applied to forward di-hadron production at RHIC. However, it was realized later
that formula (\ref{dijet}) must be also supplemented with Sudakov-type factors~\cite{Mueller:2012uf,Mueller:2013wwa},
in order to consistently resum the large logarithms that emerge when $Q_s\ll|{\bf p_{t1}}|,|{\bf p_{t2}}|.$

These are tasks that we leave for future work. In this study, we shall instead investigate forward dijets using 
 the HEF formalism, which we briefly recall below.

\subsection{The HEF framework}

Double-inclusive particle production in the HEF is obtained from the following factorization formula:
\begin{equation}
  \frac{d\sigma^{pA\rightarrow {\rm dijets}+X}}{dy_1dy_2d^2p_{1t}d^2p_{2t}} 
  =
  \sum_{a,c,d} 
  \frac{1}{16\pi^3 (x_1x_2 S)^2}
  |\overline{{\cal M}_{ag\to cd}}|^2
  x_1 f_{a/p}(x_1,\mu^2)\,
  {\cal F}_{A}(x_2,|{\bf p_{1t}}+{\bf p_{2t}}|)\frac{1}{1+\delta_{cd}}\ .
  \label{eq:cs-main}
\end{equation}
It is an extension of the collinear-factorization formulation, with a transverse
momentum dependent gluon distribution for the nucleus (or proton in the p+p
case) probed at small $x$. That distribution is simply related to $\tilde{N}_F$
by:
\begin{equation}
{\cal F}_A(x,k)=\frac{N_c}{\alpha_s (2\pi)^3}\int d^2b\int d^2r\
e^{-i{\bf k}\cdot{\bf r}}\nabla^2_{r}\ N_F(x,r,{\bf b})
=\frac{N_c\ k^2}{2\pi\alpha_s}\tilde{N}_F(x,k)\ .
\label{eq:transf0}
\end{equation}
The quantities 
$|\overline{{\cal M}_{ag\to cd}}|^2$ are $2\to 2$ polarization-averaged matrix elements with an off-shell small-$x$ gluon. The following partonic sub-processes contribute to the production of the dijet system:
\begin{equation}
  qg  \to  qg\,,
  \qquad \qquad 
  gg  \to  q\bar q\,,
  \qquad \qquad 
  gg  \to  gg\,.
\end{equation}
In contrast to formula (\ref{dijet}), the large $N_c$ limit is not assumed here,
hence the $gg\to q\bar q$ sub-process is not neglected. The corresponding amplitudes were computed in \cite{Deak:2009xt} and
cross-checked independently in \cite{vanHameren:2012uj,vanHameren:2012if}, using
different methods. The expressions are given in Appendix~\ref{sec:appa}.

Finally, let us emphasize that this framework should be considered as a model,
since, in general, there exists no transverse momentum factorization
 theorem 
 for jet
production in hadron-hadron collisions. Even in the nearly back-to-back limit
$|{\bf p_{t1}}+{\bf p_{t2}}|\ll|{\bf p_{t1}}|,|{\bf p_{t2}}|$, where the
factorization \eqref{dijet} could be established for dilute/dense collisions
(p+A or forward production), several gluon distributions are involved which is
not the case in \eqref{eq:cs-main}. We do note however that there exist a
kinematic window, namely $Q_s\ll|{\bf p_{t1}}+{\bf p_{t2}}|\ll|{\bf
p_{t1}}|,|{\bf p_{t2}}|$, in which that formula can be motivated \cite{Iancu:2013dta}.
Indeed, when $|{\bf p_{t1}}+{\bf p_{t2}}|\ll|{\bf p_{t1}}|,|{\bf p_{t2}}|$, the
off-shell matrix elements
given in Appendix~\ref{sec:appa} reduce to those of~Eq.~\eqref{dijet}, and when
$Q_s\ll|{\bf p_{t1}}+{\bf p_{t2}}|$, the different gluon distribution of
Eq.~(\ref{dijet}) have the same asymptotic behaviour as $\tilde{N}_F$. We shall elaborate on this in a future publication.

\section{Non-linear evolution of the unintegrated gluon distributions}

In order to complete our formulation of the forward jet cross sections, we discuss now the $x$ evolution of the unintegrated gluon distributions.

\subsection{The rcBK evolution}

In the CGC framework, the evolution of $\tilde{N}_F$ is obtained from the evolution of the dipole scattering amplitude $N_F(x,r,{\bf b})$ (see (\ref{phihyb})), with the assumption that the impact parameter dependence of $N_F$ factorizes, and therefore does not mix with the evolution.

The evolution equation of the dipole amplitude, known as the Balitsky-Kovchegov equation \cite{Balitsky:1995ub,Kovchegov:1999yj}, supplemented with running coupling corrections (henceforth referred to as rcBK equation) reads ($r_i=|\bf r_i|$)
\begin{eqnarray}
  \frac{\partial N_{F}(r,x)}{\partial\ln(x_0/x)}=\int d^2r_1\
  K^{{\rm run}}({\bf r},{\bf r_1},{\bf r_2})
  \left[N_{F}(r_1,x)+N_{F}(r_2,x)
-N_{F}(r,x)\right.\nonumber\\
\left. -N_{F}(r_1,x)\,N_{F}(r_2,x)\right]\ ,
\label{bk1}
\end{eqnarray}
with ${\bf r_2}\equiv{\bf r}-{\bf r_1}$ and where $x_0$ is some initial value
for the evolution (usually chosen to be 0.01). ${K}^{\rm run}$ is the evolution
kernel including running coupling corrections. Different prescriptions have been
proposed in the literature for $K^{\rm run}$. As shown in \cite{Albacete:2007yr}, Balitsky's prescription minimizes the role of
higher {\it conformal} corrections:
\begin{equation}
  K^{{\rm run}}({\bf r},{\bf r_1},{\bf r_2})=\frac{N_c\,\alpha_s(r^2)}{2\pi^2}
  \left[\frac{1}{r_1^2}\left(\frac{\alpha_s(r_1^2)}{\alpha_s(r_2^2)}-1\right)+
    \frac{r^2}{r_1^2\,r_2^2}+\frac{1}{r_2^2}\left(\frac{\alpha_s(r_2^2)}{\alpha_s(r_1^2)}-1\right) \right]\,.
\label{kbal}
\end{equation}

The rcBK evolution is independent of whether the target is a proton or a nucleus. That difference is accounted for in the initial condition. The parametrization usually used is
\begin{equation}
N_F(r,x\!=\!x_0)=
1-\exp\left[-\frac{\left(r^2\,Q_{s0}^2\right)^{\gamma}}{4}\,
  \ln\left(\frac{1}{\Lambda\,r}+e\right)\right]\ ,
\label{ic}
\end{equation}
where, $\Lambda=0.241$~GeV, $Q_{s0}$ is the saturation scale at the initial $x_{0}$ and $\gamma$ is a dimensionless parameter
that controls the steepness of the unintegrated gluon distribution for momenta above the initial saturation scale $k_t>Q_{s0}$. In the proton case,
the free parameters are obtained from a fit \cite{Albacete:2010sy} to HERA proton structure function data \cite{Aaron:2009aa}: $\gamma=1.119$ and $Q_{s0}^2=0.168$ GeV$^2$. 
In the nucleus case, we will use identical parameters except for $Q^2_{s0}$ 
 for which we shall use
\begin{equation}
Q_{s0}^{2,A}=d\, Q_{s0}^2
\label{eq:QA}
\end{equation}
and vary the $d$ parameter between 2 and 4. 
We note that the resulting unintegrated gluon distributions $\tilde{N}_F$ are those used in the rcBK Monte Carlo \cite{oai:arXiv.org:1209.2001}.

The corresponding BK equation in momentum space, for the unintegrated gluon
density ${\cal F}$, reads \cite{Kutak:2003bd,Nikolaev:2006za}:
\begin{multline} 
{\cal F}_p(x,k^2) \; = \; {\cal F}^{(0)}_p(x,k^2) \\
+\,\frac{\alpha_s N_c}{\pi}\int_x^1 \frac{dz}{z} \int_{k_0^2}^{\infty}
\frac{dl^2}{l^2} \,   \bigg\{ \, \frac{l^2{\cal F}_p(\frac{x}{z},l^2)\,   -\,
k^2{\cal F}_p(\frac{x}{z},k^2)}{|l^2-k^2|}   +\,
\frac{k^2{\cal F}_p(\frac{x}{z},k^2)}{|4l^4+k^4|^{\frac{1}{2}}} \,
\bigg\} \\
-\frac{2\alpha_s^2}{R^2}\left[\left(\int_{k^2}^{\infty}\frac{dl^2}{l^2}{\cal F}_p(x,l^2)\right)^2
+
{\cal F}_p(x,k^2)\int_{k^2}^{\infty}\frac{dl^2}{l^2}\ln\left(\frac{l^2}{k^2}\right){\cal F}_p(x,l^2)\right]\ .
\label{eq:fkovres0} 
\end{multline}
In this formulation one can relatively easily include dominant corrections of higher orders as is discussed in the next section.
Note that, in order to write the non-linear term of this equation for the impact-parameter-integrated gluon distribution, we have assumed that the impact parameter integral gives $\int d^2b=\pi R^2$ where $R$ is a radius of the target proton. The equation for ${\cal F}_A$ in the nuclear case is discussed below.

\subsection{The KS gluon density}

In principle, the gluon density in the HEF framework ${\cal F}_p$ evolves with $x$ according to the Balitsky-Fadin-Kuraev-Lipatov (BFKL) evolution \cite{bfkl}. However, the 
 Ciafaloni-Catani-Fiorani-Marchesini  (CCFM)
 equation \cite{Ciafaloni:1987ur,Catani:1989sg,Catani:1989yc} can also be used to take into account coherence effects in the evolution. The coherence effect in the emissions of gluons is a manifestation of their quantum nature. Its inclusion in the evolution leads to angular ordering of subsequently emitted gluons during evolution. If the newly emitted gluon violates the imposed ordering it does not contribute to building up the gluon density. Consequently  emitted gluons which build up the density have to be emitted with increasing angle. The coherence will also introduce a constraint on maximal allowed angle in the evolution of gluons which is linked to the transverse momentum of the measured dijet system. 

The BK evolution introduced above can also be used to take into account non-linear corrections to the BFKL evolution. Both non-linear and coherence effects can be included simultaneously \cite{Kutak:2011fu,Kutak:2012qk,Kutak:2012yr}, which is in principle required, since we are interested in observables sensitive to both the saturation regime and high-$p_t$ physics. We will be able to do so when the resulting gluon density constrained by HERA data becomes available. So far, only purely theoretical, numerical results are available~\cite{Deak:2012mx,Kutak:2013yga}.  

In the meantime, we shall use the simplified equation proposed in \cite{Kutak:2003bd,Kutak:2004ym}:
\begin{multline} 
{\cal F}_p(x,k^2) \; = \; {\cal F}^{(0)}_p(x,k^2) \\
+\,\frac{\alpha_s(k^2) N_c}{\pi}\int_x^1 \frac{dz}{z} \int_{k_0^2}^{\infty}
\frac{dl^2}{l^2} \,   \bigg\{ \, \frac{l^2{\cal F}_p(\frac{x}{z},l^2) \, \theta(\frac{k^2}{z}-l^2)\,   -\,
k^2{\cal F}_p(\frac{x}{z},k^2)}{|l^2-k^2|}   +\,
\frac{k^2{\cal F}_p(\frac{x}{z},k^2)}{|4l^4+k^4|^{\frac{1}{2}}} \,
\bigg\} \\  + \, \frac{\alpha_s (k^2)}{2\pi k^2} \int_x^1 dz \,
\Bigg[\left(P_{gg}(z)-\frac{2N_c}{z}\right) \int^{k^2}_{k_0^2} d l^{
2}\,{\cal F}_p\left(\frac{x}{z},l^2\right)+zP_{gq}(z)\Sigma\left(\frac{x}{z},k^2\right)\Bigg] \\
-\frac{2\alpha_s^2(k^2)}{R^2}\left[\left(\int_{k^2}^{\infty}\frac{dl^2}{l^2}{\cal F}_p(x,l^2)\right)^2
+
{\cal F}_p(x,k^2)\int_{k^2}^{\infty}\frac{dl^2}{l^2}\ln\left(\frac{l^2}{k^2}\right){\cal F}_p(x,l^2)\right]\,,
\label{eq:fkovresKS} 
\end{multline} 
where $z=x/x'$. This is the BK equation extended to take into account higher-order corrections
coming from including non-singular pieces of the gluon splitting function, kinematic constraint effects
and contributions from sea quarks. The input gluon distribution ${\cal F}^{(0)}_p(x,k^2)$ is is given by 
\begin{equation}
 {\cal F}^{(0)}_p(x,k^2)=\frac{\alpha_S(k^2)}{2\pi k^2}\int_x^1
 dzP_{gg}(z)\frac{x}{z}g\left(\frac{x}{z},k_0^2\right)\,,
 \label{eq:initial-cond} 
\end{equation}
where $xg(x,k_0^2)$ is the integrated gluon distribution at the initial scale,
which we set to $k_0^2=1\, \text{GeV}^2$. We take the following parametrization
\begin{equation}
 xg(x,1\, {\rm GeV}^2)=N(1-x)^{\beta} (1-D x)\,.
 \label{eq:input-param} 
\end{equation}
The parameters $N$, $\beta$, $D$, together with the proton radius, $R$, were
constrained with a fit to HERA data \cite{Aaron:2009aa} in \cite{Kutak:2012rf},
hence we will refer to the resulting gluon density as the ``KS gluon''. The fit gave the
following result:  $N=0.994$, $\beta= 18.6$, $D=-82.1$, $R=2.40\,\text{GeV}^{-1}$. 

In order to use Eq.~(\ref{eq:fkovresKS}) to obtain the nuclear gluon density
${\cal F}_A$, one needs to make the following formal substitution
\begin{equation}
\frac{1}{R^2} \to  c\, \frac{A}{R_A^2}\,,
\qquad {\rm where} \quad
R^2_{\text{A}}= R^2\, A^{2/3} \,.
\label{eq:radius}
\end{equation}
In the above equation, 
$R_A$ is the nuclear radius,
where $A$ is the mass number ($A=207$ for Pb), and $c$ is a parameter that we
shall vary between 0.5 and 1 to assess the uncertainty related to the nonlinear
term. The density ${\cal F}_A$ obtained from Eq.~(\ref{eq:fkovresKS}) with the
substitution above is the nuclear gluon density normalized to the number of nucleons in the nuclei. 

Since the KS evolution equation is already $A$-dependent through the non-linear term (it has to be so since ${\cal F}_A$ is an impact parameter integrated distribution), our prescription for the initial condition is to choose the same in the nuclear case as  in the proton case ${\cal F}^{(0)}_A(x,k^2)={\cal F}^{(0)}_p(x,k^2)$. This is an interesting difference with the rcBK gluon density (for which it is the initial condition that is A-dependent while the evolution equation is A-independent), the impact of which we shall discuss in the following section.

\section{Forward-forward dijet production at the LHC}

\begin{figure}[t]
  \begin{center}
    \includegraphics[width=0.32\textwidth]{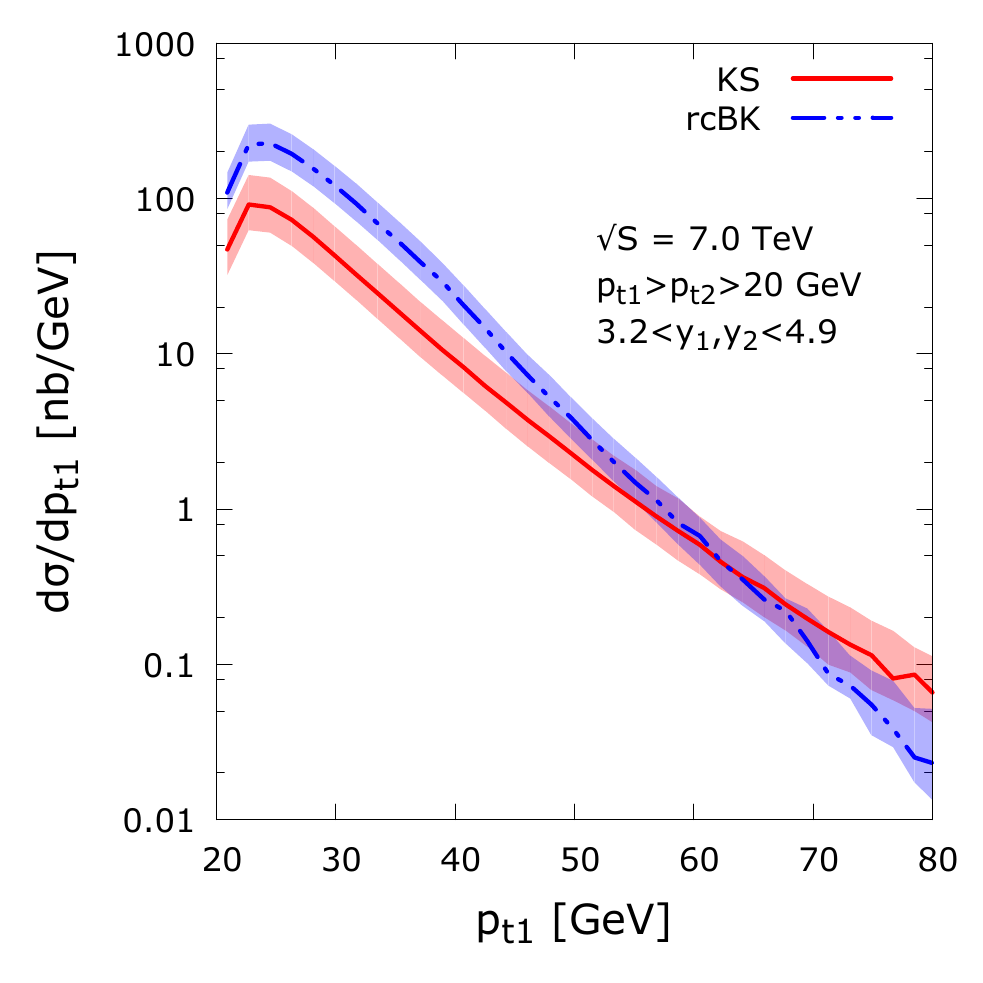}
    \includegraphics[width=0.32\textwidth]{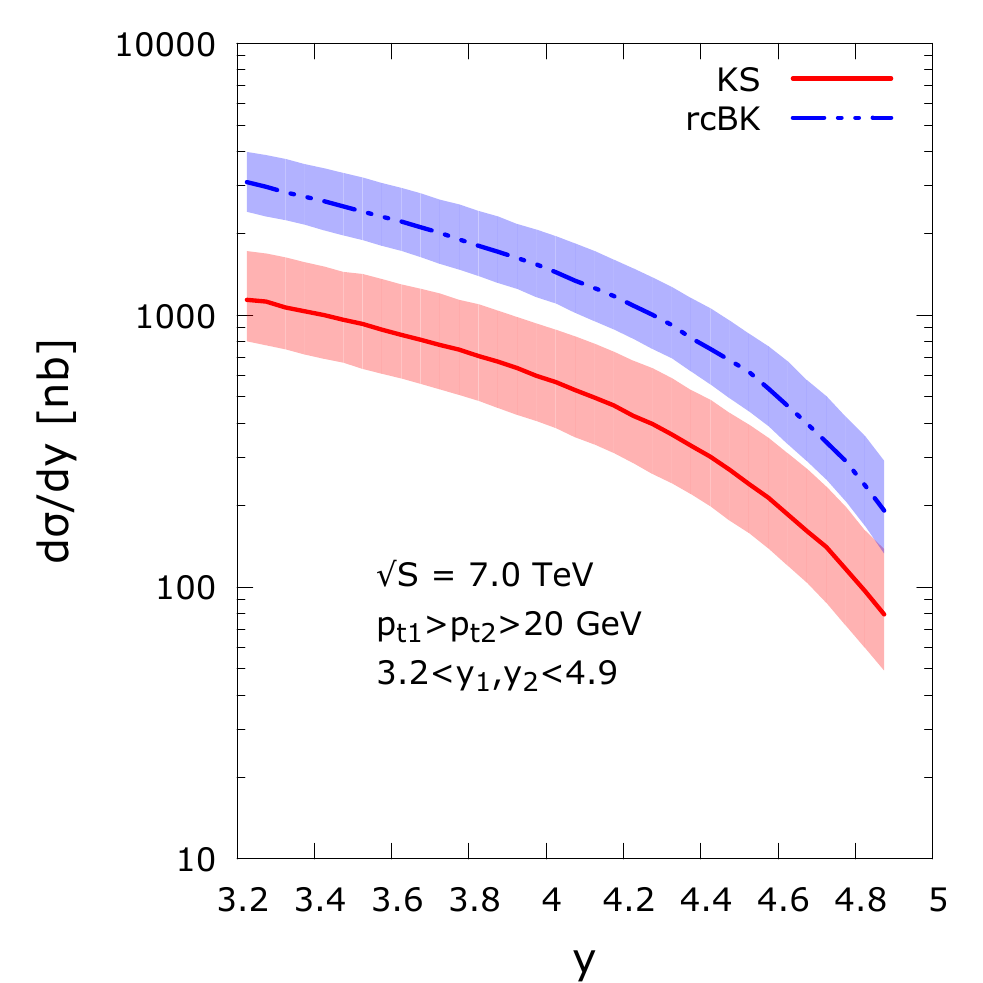}
    \includegraphics[width=0.32\textwidth]{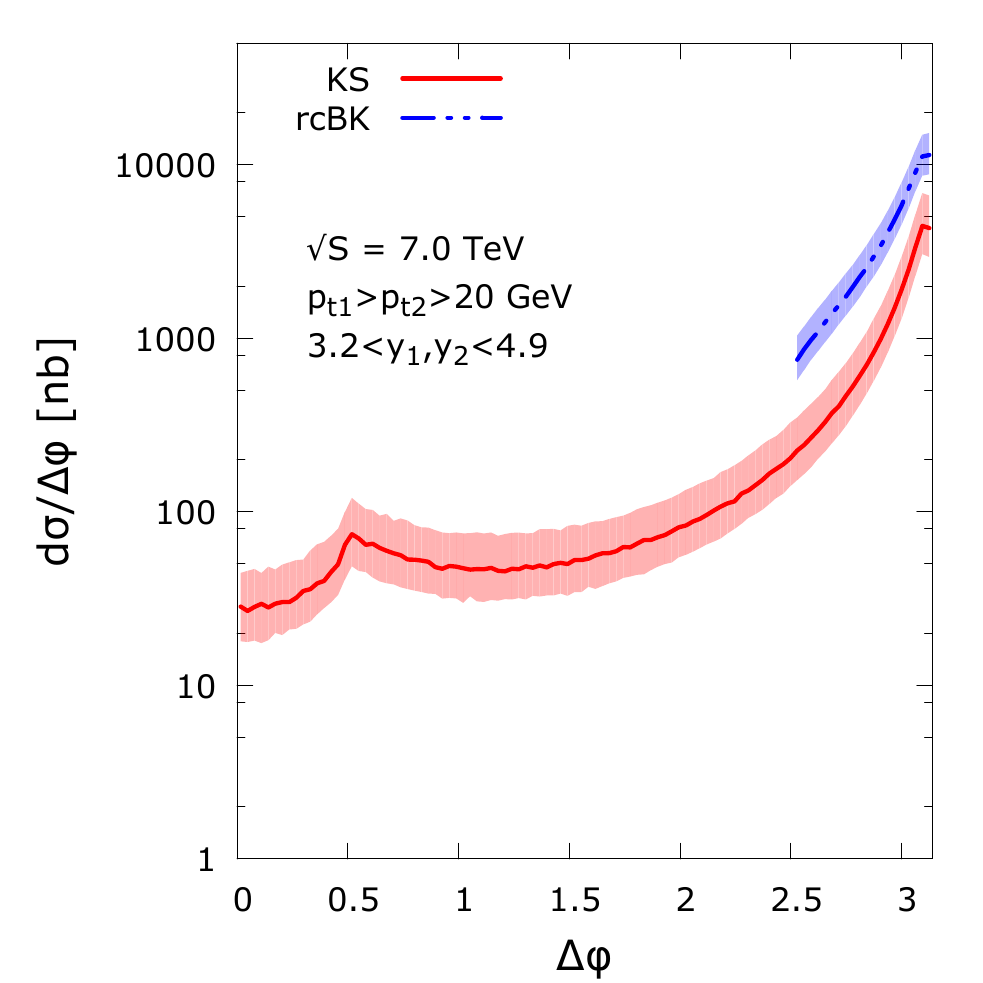}
  \end{center}
  \caption{
  \small
  Differential cross sections for forward-forward jet productions in p+p
  collisions, as functions of transverse momentum of the leading jet (left),
  jet rapidity (middle), and the azimuthal angle between the two hardest jets. 
  The two bands correspond to two different unintegrated gluon distributions
  used for calculations. The width of the bands comes from varying the
  renormalization and factorizations scales by factors $\frac12$ and 2 around the
  central value taken as the average $p_t$ of the two leading jets.
  }
  \label{fig:distributions}
\end{figure}

We move now to the numerical results for forward dijet production at the LHC in
p+p and p+Pb collisions.
 
Our predictions were generated with the forward region defined as the rapidity range
$3.2 < y < 4.9$. The two hardest jets
 (sorted according to their $p_t$, $p_{t1}>p_{t2}$) 
 are required to lie within this region.
In order to cut-off collinear and soft singularities, we use the anti-$k_t$
algorithm~\cite{Cacciari:2008gp}, with radius $R=0.5$, and we require each jet to
have $p_t$ above 20~GeV. Since, in our modelling, a jet is a single parton, the
application of the anti-$k_t$ jet algorithm boils down to a cut in the $y-\phi$
plane. If the distance between two partons $\Delta R_{ij}=
\sqrt{\Delta\phi_{ij}^2+\Delta y_{ij}} > R$ then  they form two separate jets,
otherwise they form one jet and the corresponding event is rejected.

The calculations were performed and cross-checked using three independent Monte Carlo
programs: 
(i) \verb+forward+~\cite{forward}, which is a direct implementation of
Eq.~(\ref{eq:cs-main}) and uses Vegas
algorithm~\cite{Lepage:1977sw,Lepage:1980dq} for integration, 
(ii) a program implementing the method of \cite{vanHameren:2012if} using Kaleu~\cite{vanHameren:2010gg} and Parni~\cite{vanHameren:2007pt} for integration, and  
(iii) $\mathtt{LxJet}$ \cite{LxJet:2013} using the FOAM
algorithm~\cite{Jadach:2002kn} for integration.
As mentioned in the previous section, we performed our computations using two
different unintegrated gluon distributions, rcBK and KS.
For the collinear PDFs, which also enter the HEF formula~(\ref{eq:cs-main}),
we took the general-purpose CT10 set~\cite{Lai:2010vv}.
For the central value of the factorization and renormalization scale, we chose
the average transverse momentum of the two leading jets, $\mu_F=\mu_R=
\frac12 (p_{t1}+p_{t2})$.

Fig.~\ref{fig:distributions} shows the differential cross sections for p+p
collision at the center of mass energy of 7 TeV. Two bands come from using
two different unintegrated gluons: rcBK and KS.
The width of a band corresponds to the renormalization and factorization
scale uncertainties obtained by varying the central value by factors 2 and
$\frac12$.
In our framework, scale dependence enters the collinear PDFs and the strong
coupling constant, $\alpha_s$, that resides inside the matrix element. 
Since our calculation is formally leading order, in terms of powers of
$\alpha_s$, there is nothing to compensate for the scale dependence, thus the
uncertainties for the absolute predictions, shown in
Fig.~\ref{fig:distributions}, are substantial. They will cancel to large extent for the
nuclear modification ratios, which we discuss later in this section.

The results with the KS gluon~\cite{Kutak:2012rf} are absolute predictions both in
terms of shape and normalization. For the rcBK case, however, since the original
gluon density is obtained prior to impact-parameter integration, while the HEF
formula~(\ref{eq:cs-main}) requires an impact-parameter integrated gluon distribution, there is an
ambiguity as to which normalization should be chosen. In the rcBK bands in
Fig.~\ref{fig:distributions}, we have adjusted the normalization so that it gives
a cross section comparable to the KS case, in the range $p_{t1}=50-80$ GeV, a window where
the KS predictions reproduce  the data well in the central-forward case \cite{Kutak:2012rf}.
We emphasize however that when comparing the rcBK and KS distribution, one should concentrate rather on the
shape differences. This ambiguity will be removed later in this section, when we turn to nuclear modification factors, which are
ratios of distribution where the normalization factors cancel.

In the left plot from Fig.~\ref{fig:distributions}, we show the distribution of
transverse momentum of the leading jet. We see that the differences between KS
and rcBK are not only in normalization but also in shape with the latter gluon
leading to a steeper $p_{t}$ spectrum.
The middle plot compares rapidity distributions in the two scenarios. Here, the
difference is mostly in the normalization while the shapes are very similar.
The right plot in Fig.~\ref{fig:distributions} shows an observable which is particularly well suited
to study saturation, the azimuthal decorrelations between two hardest jets. This
observable is nothing else but the azimuthal distance
$\Delta\phi=\phi_1-\phi_2$. If we look at Eq.~(\ref{eq:cs-main}), we see that the
unintegrated gluon density is taken at the gluon transverse momentum
\begin{equation}
  k_{tg}^2 = |{\bf p_{1t}}+{\bf p_{2t}}|^2 = 
  p_{t1}^2 + p_{t2}^2 + 2p_{t1}p_{t2} \cos\Delta\phi\,.
\end{equation}
That implies that in the limit $\Delta\phi \to \pi$, which corresponds to almost
back-to-back dijet configurations, the gluon is probed at a very low $k_t$, where
saturation is expected to be important.
As we see in Fig.~\ref{fig:distributions} (right) the two gluons lead to a
somewhat different shape of the decorrelation spectrum above $\Delta \phi\sim 2.5$.
 We do not plot the rcBK curve 
 below 
 $\Delta \phi \sim 2.5$,
since $|{\bf p_{t1}}+{\bf p_{t2}}|$ is becoming too large compared to the saturation scale
 where the model does not apply.
The peak in the KS result around $\Delta \phi \sim 0.5$ and the rapid decrease
below that value comes from using the anti-$k_t$ jet algorithm with $R=0.5$,
which leads to significant depletion of the cross section in this region.

\begin{figure}[t]
  \begin{center}
    \includegraphics[width=0.45\textwidth]{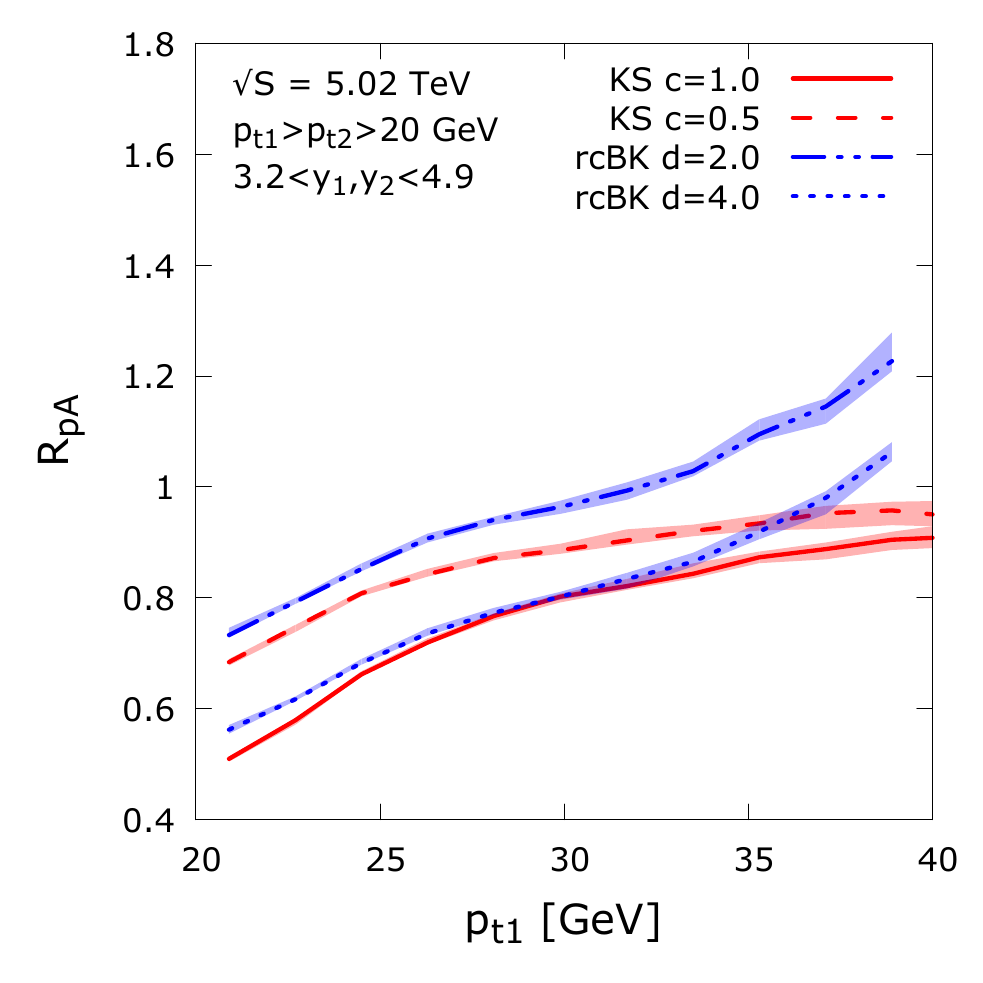}
    \includegraphics[width=0.45\textwidth]{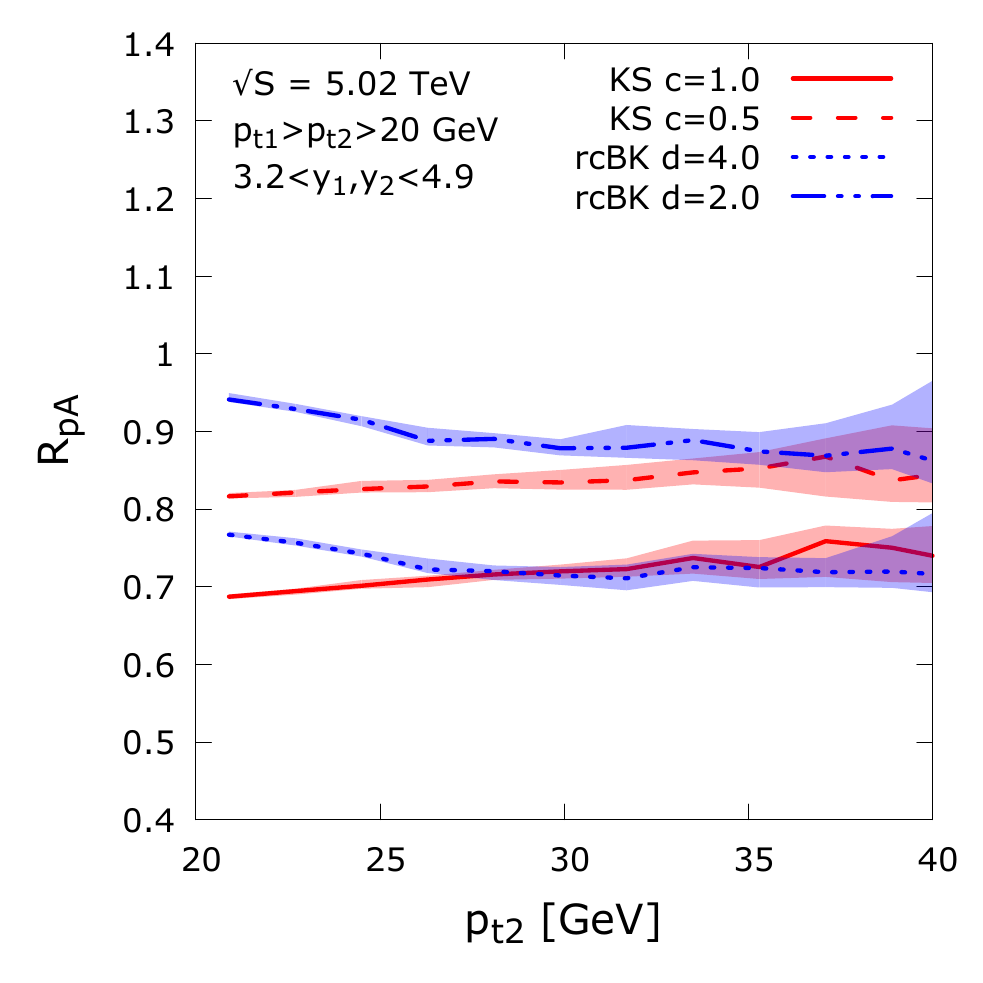}
  \end{center}
  \caption{
  \small
  Nuclear modification ratios, defined in Eq.~(\ref{eq:RpA}), as
  functions of $p_t$s of the leading~(left) and subleading~(right) jets produced
  in the forward region.
  Different bands correspond to different unintegrated gluon distributions (KS
  and rcBK) used for calculations. To asses the uncertainty related to the
  nonlinear effects, the KS result was computed with two values of the $c$
  parameter, defined in Eq.~(\ref{eq:radius}), and the rcBK prediction was
  obtained with two values of the $d$ parameter from Eq.~(\ref{eq:QA}).
  The width of the bands comes from varying the
  renormalization and factorizations scales by factors $\frac12$ and 2 around the
  central value taken as the average $p_t$ of the two leading jets.
  }
  \label{fig:RpA1}
\end{figure}

The overall conclusion one can draw from the results shown in
Fig.~\ref{fig:distributions} is that two different unintegrated gluons, which
describe equally well the inclusive $F_2$~\cite{Kutak:2012rf,Albacete:2010sy}, can lead to different shapes
of the experimentally relevant distributions in the forward-forward dijet
production. This may be an indication of the importance of higher order
corrections in the evolution of the unintegrated gluon density present in the KS
case.

\begin{figure}[t]
  \begin{center}
    \includegraphics[width=0.45\textwidth]{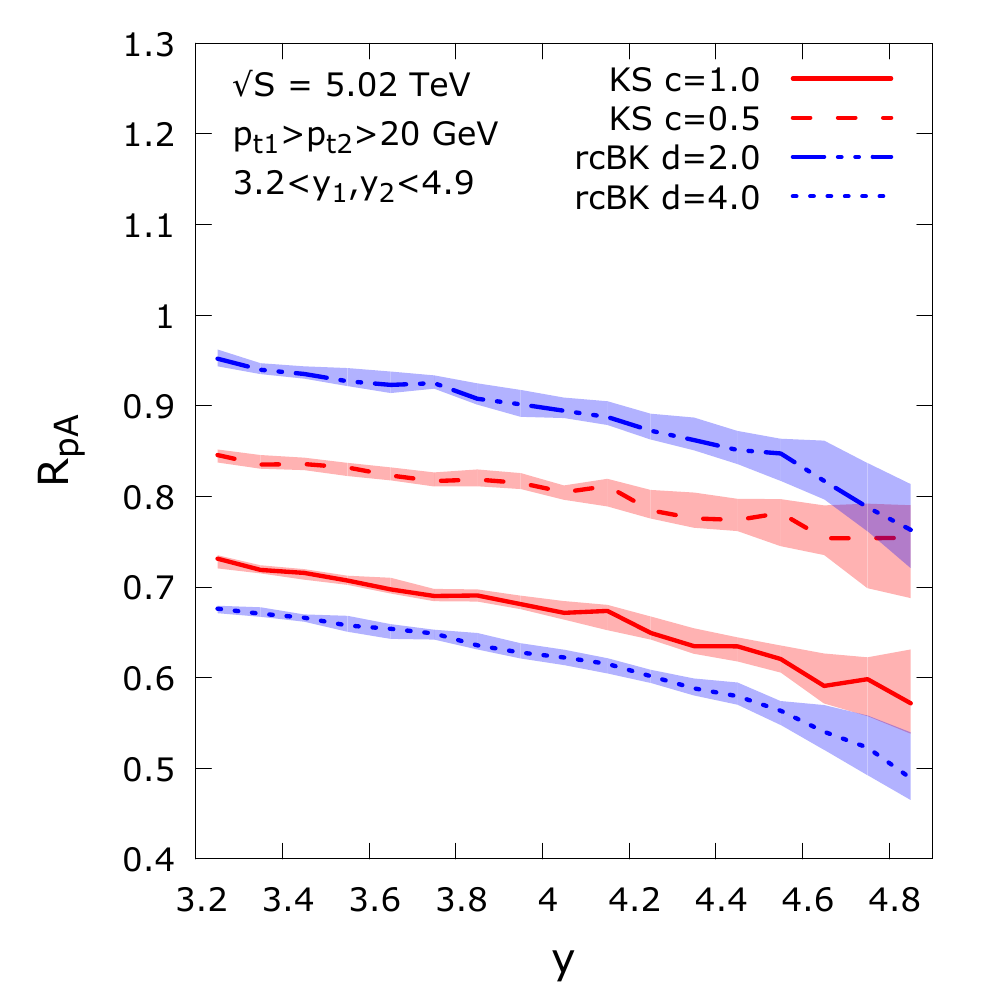}
    \includegraphics[width=0.45\textwidth]{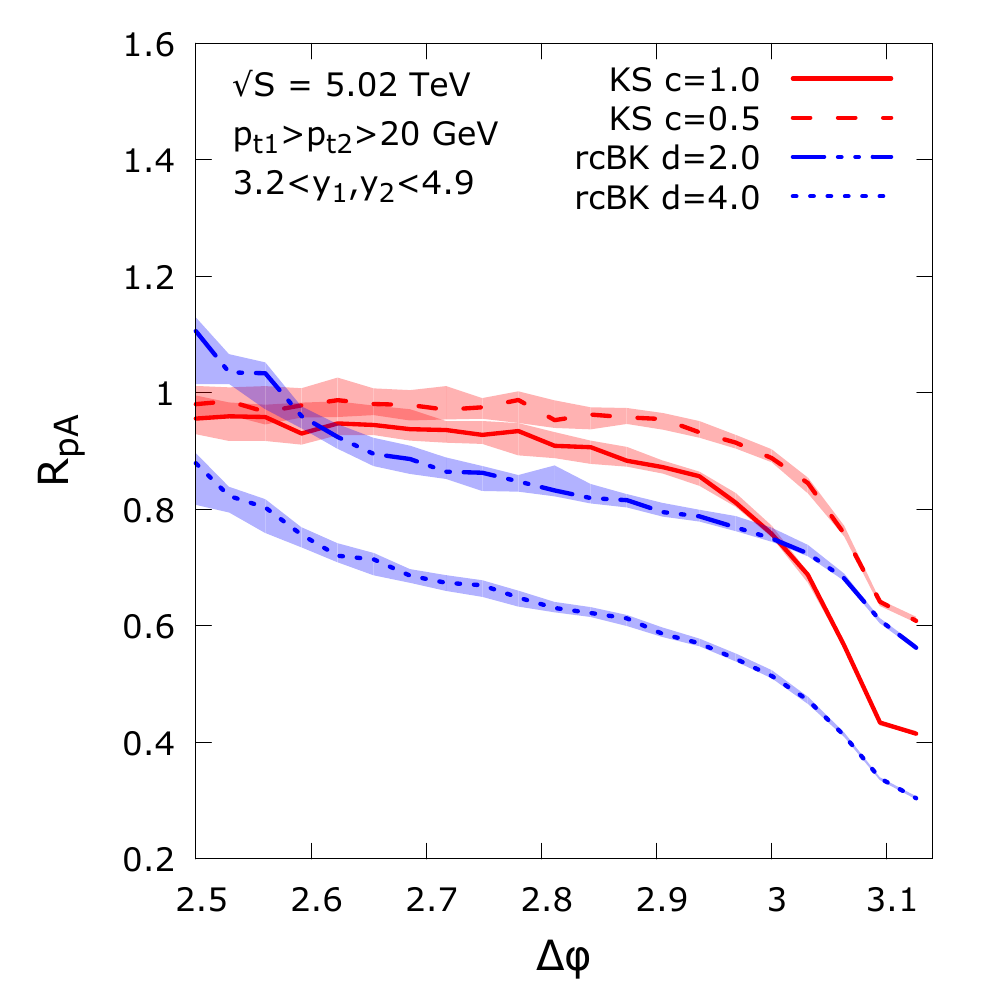}
  \end{center}
  \caption{
  \small
  Nuclear modification ratios, defined in Eq.~(\ref{eq:RpA}), as
  functions of jet rapidity (left) and the azimuthal distance between
  two hardest jets produced in the forward region~(right).
  All details as in Fig.~\ref{fig:RpA1}.
  }
  \label{fig:RpA2}
\end{figure}

Let us turn to the discussion of possible signatures of saturation. For that we
shall look at the forward-forward dijet production in p+A collisions and compare
to the previously described p+p case.
In Figs.~\ref{fig:RpA1} and \ref{fig:RpA2} we show the nuclear  modification
factors defined for each observable ${\cal O}$ as
\begin{equation}
R_{\rm pA} = \frac{\displaystyle \frac{d\sigma^{p+A}}{d{\cal O}}}
                 {\displaystyle A\ \frac{d\sigma^{p+p}}{d{\cal O}}}\,.
\label{eq:RpA}
\end{equation}
If the case of absence of nonlinear effects or in the case in which they are
equally strong in the nucleus and in the proton, this ratio equals 1. If,
however, the nonlinear evolution plays a more important role in the case of the
nucleus, the $R_{\rm pA}$ ratio will be suppressed below 1.
 
The plots in Fig.~\ref{fig:RpA1} show the $R_{\rm pA}$ ratios for the $p_t$ of
the leading~(left) and subleading~(right) jet. In Fig.~\ref{fig:RpA2} we have
similar ratios for rapidity and azimuthal angle distributions.
For each gluon we consider two scenarios to assess possible uncertainties of our
prediction. In the rcBK case, we use 
 $d=2.0$ and $d=4.0$ (c.f. Eq.~(\ref{eq:QA})), 
while in the KS case, we use $c=0.5$ and $c=1.0$ (see formula (\ref{eq:radius})).
As we see in the plots, the scale uncertainty is greatly reduced compared to the
distributions shown in Fig.~\ref{fig:distributions}. The qualitative behaviour
of the predictions with two gluons is very similar. They differ mostly at the
quantitative level.
In particular, in the case of $p_t$ of the leading jet, we observe a suppression of the order
of $20-30\%$  at low $p_t$ for rcBK and $30-50\%$ for the KS gluon.
For the subleading jet this suppression is smaller in both cases. The rapidity
$R_{\rm pA}$ ratios are also significantly below 1, especially in the very forward
region, which corresponds to probing the unintegrated gluon at low $x$, hence in
the domain with strong sensitivity to saturation effects.
Finally, in the case of decorrelations, $\Delta \phi$, both gluons lead to even
up to $60\%$ suppression in the back-to-back limit $\Delta \phi \to \pi$.

\begin{figure}[t]
  \begin{center}
    \includegraphics[width=0.45\textwidth]{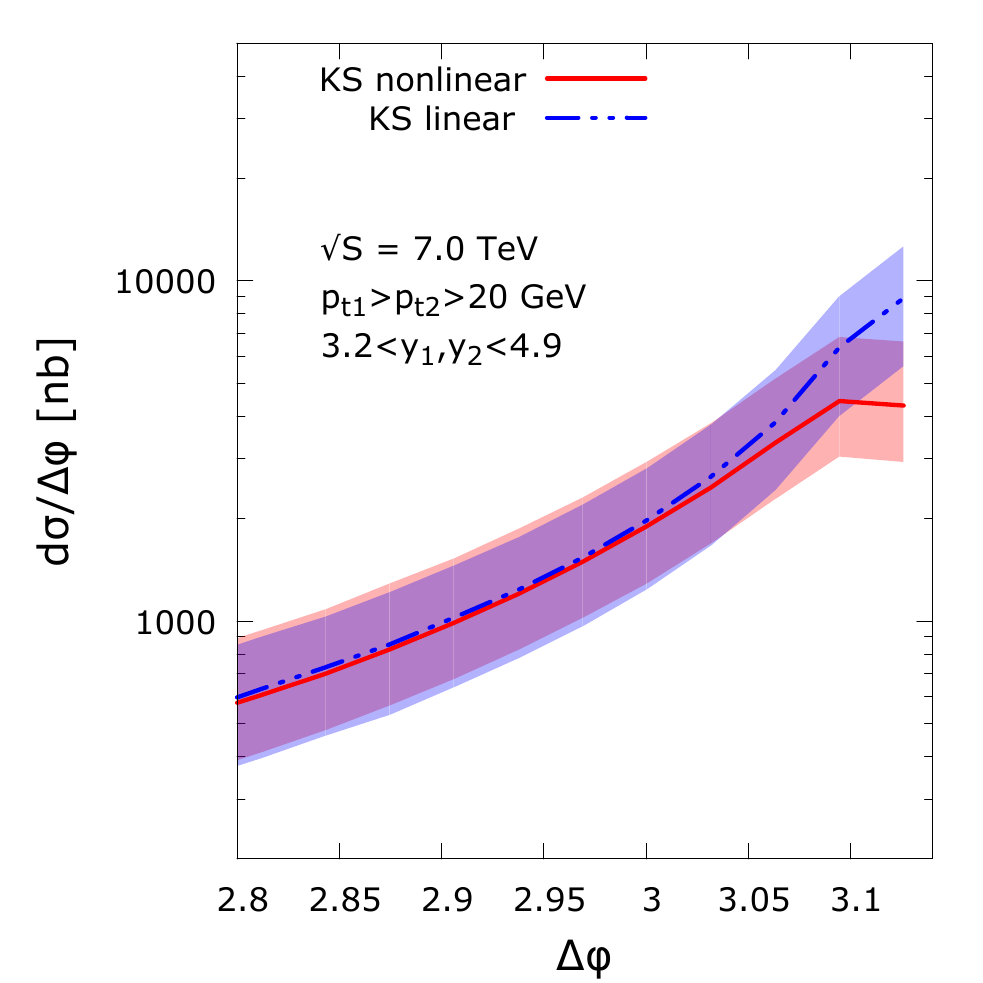}
    \includegraphics[width=0.45\textwidth]{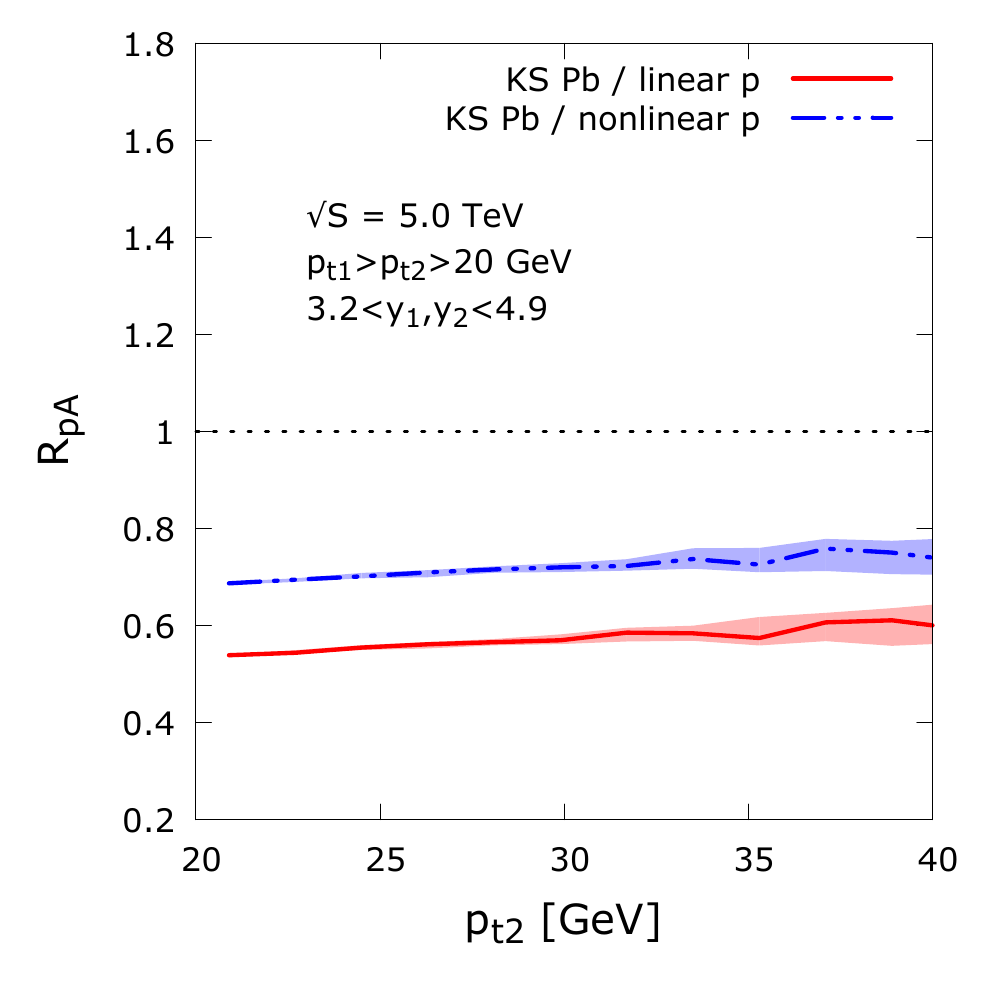}
  \end{center}
  \caption{
  \small
  Comparisons of predictions obtained with a linear vs non-linear proton evolution,
  for the differential cross section in p+p collisions as a function of the azimuthal angle (left),
  and the nuclear modification factor as a function of $p_t$ of the subleading jet.
  KS gluon densities are used, with parameter $c=1$ in the nuclear case.
  }
  \label{fig:linear}
\end{figure}

To finish, we illustrate in Fig.~\ref{fig:linear} the impact of saturation effects in the KS
evolution by switching off the non-linear term in the evolution. We note that the parameters
of this alternative gluon distribution for the proton, obtained with linear evolution, are
re-adjusted in order to keep a good description of DIS data from HERA. 
The left plot shows the impact of non-linear effects on the differential cross section in p+p
collisions as a function of the azimuthal angle, and it is large, as expected, near $\Delta\Phi=\pi$.
The right plot shows, in the case of the sub-leading jet $p_t$ dependence, by how much $R_{pPb}$
is reduced if the nuclear gluon density Lead is still subject to non-linear effects, but not proton one.
Of course with the KS gluon distributions, by construction, if non-linear evolution is switched
off both in the proton and nuclear cases, then $R_{pPb}=1.$

\section{Summary}

In this paper we studied forward-forward dijet production and we argued that this
process is particularly attractive from low-$x$ point of view. Using the High Energy
Factorization approach (Eq.~(\ref{eq:cs-main})), we provided predictions for 
distributions and nuclear modification factors in p+Pb vs p+p collisions, as
functions of the transverse momenta, rapidities and relative azimuthal distance
of the two hardest jets produced in the forward the region.

Let us first recall that in a small corner of the phase space, for nearly back-to-back di-jets
($|{\bf p_{t1}} + {\bf p_{t2}}| \leq Q_s$), our calculations should be improved by implementing
the more complete factorization formula in Eq.~(\ref{dijet}). Even if non-linear effects are the
biggest in there, this small limitation does not affect our conclusions since sizable
saturation effects are also seen outside of that kinematic window.

To compute our predictions, we used gluon distributions from two different extensions
of the BK equation: the rcBK gluon density obtained from Eq.~(\ref{bk1}), and the KS
gluon density obtained from Eq.~(\ref{eq:fkovresKS}). We found that both lead to a very similar
behavior of the nuclear modification factors in the domain where the two extensions
are applicable. In particular, both sets of predictions suggest significant effects of gluon
saturation as one goes from p+p to p+Pb.

Even though it was recently realized that the outcome of high-energy proton-nucleus
collisions is quite sensitive to the fact that the nucleon positions in the nucleus fluctuate
event by event, so far our $R_{pPb}$ predictions have been obtained using an impact-parameter
averaged nuclear saturation scale: Eq.~(\ref{eq:QA}) in the rcBK case and more indirectly from
Eq.~(\ref{eq:radius}) in the KS case. However, in order to estimate the corresponding
uncertainty, we have varied the nuclear saturation strength parameter $c$ (KS) or $d$ (rcBK).
A more complete study should certainly include such nucleon-level fluctuation effects,
nevertheless our present results are enough to motivate experimental measurements at the LHC.

Finally, our results also show the importance of higher-order corrections to the BK equation.
They allow, for example, the extension of the applicability of the unintegrated gluon density
towards larger values of transverse momenta, which has consequences for several observables,
like the $\Delta \phi$ distribution away from $\pi$. In the equations used in this work those
corrections were implemented as educated guesses, but in principle they deserve a rigorous derivation.

\section*{Acknowledgements}
The work of A. van Hameren, K. Kutak and P. Kotko has been supported by Narodowe
Centrum Badan i Rozwoju with grant LIDER/02/35/L-2/10/NCBiR/2011.
KK and CM acknowledge the support of the European Community under the FP7
"Capacities" Programme in the area of Research Infrastructure, as this work was
initiated during the h$_3$QCD ECT* workshop.

\appendix
\section{Appendix}
\label{sec:appa}

For convenience, we collect here the matrix elements for the $2\rightarrow 2$
partonic processes, computed in \cite{Deak:2009xt, vanHameren:2012uj,vanHameren:2012if}, and expressed in terms of the transverse momentum and rapidity of
the final the state partons. 
Parametrizing the final state momenta as
\begin{equation}
p^{\mu}_i=(p_{ti}\cosh y_i,{\bf p_{ti}},p_{ti}\sinh y_i),\,\,\, i=1,2\,,
\end{equation}
we have

\begin{equation}
  \overline{|{\cal M}_{qg\rightarrow qg}|}^2=
  C_1 {\cal A}_1^\abel + \overline C_1 {\cal A}_1^\noabel\,,
  \
  \overline{|{\cal M}_{gg\rightarrow q\bar q}|}^2=
  C_2 {\cal A}_2^\abel + \overline C_2 {\cal A}_2^\noabel\,,
  \
  \overline{|{\cal M}_{gg\rightarrow g\bar g}|}^2=
  C_3 {\cal A}_3\,, 
\end{equation}
with the abelian \abel\ and nonabelian \noabel\ contributions given by
{\allowdisplaybreaks
\begin{align}
   \label{eq:A1ab}
  {\cal A}_1^\abel & =
  \frac{\left(K+e^Y\right)^2\left(\left(K+e^{-Y}\right)^2+K^2\right)}
       {2 K\left(K e^Y+1\right)\left(\cosh Y -\cos\phi \right)}\,,
  \\[10pt] 
  {\cal A}_1^\noabel & =
  2  e^Y (e^Y - \cos\phi)\, {\cal A}_1^\abel\,,
  \\[10pt]
  {\cal A}_2^\abel & =
  \frac{\left(K+e^Y\right)^2 \left(K^2 e^{Y}+e^{-Y}\right)}
       {K \left(K e^Y+1\right)^2}\,,
\end{align}
}

\begin{equation}
 {\cal A}_2^\noabel=\frac{\cos\phi}{\cosh Y - \cos\phi}\, {\cal A}_2^\abel\,,
\end{equation}

\begin{equation}
 \label{eq:A3}
 {\cal A}_3 =\frac{2\left(e^{-Y}K+1\right)^2 \left(K e^Y \left(K e^Y+1\right)+1\right)^2 (\cos \phi -2 \cosh Y)}{K^2 \left(K
   e^Y+1\right)^2 (\cos \phi -\cosh (Y))}\,,
\end{equation}
where
\begin{equation}
  \label{eq:YK}
  Y = y_1-y_2\,,
  \qquad \qquad
  K = \frac{p_{t1}}{p_{t2}}\,,
\end{equation}
and $C_1=g^4(N_c^2-1)/(2N_c^2)$, $\overline C_1=C_1C_A/(2C_F)$,
$C_2=g^4/(2N_c)$, $\overline C_2=C_2C_A/(2C_F)$, $C_3=g^4N_c^2/(N_c^2-1).$



\begin{thebibliography}{99}
\bibitem{forward}
  S.~Sapeta, $\mathtt{forward}$, the code is available on request from the author
\end{thebibliography}
\end{document}